# Understanding the Relationships between Information Architectures and Business Models: An Empirical Study on the Success Configurations of Smart Communities Based on the Internet of Things (IoT)

ZHANG Nan, School of Public Policy and Management, Tsinghua University, Beijing, China, Corresponding Author: nanzhang@tsinghua.edu.cn

ZHAO Xuejiao, School of Public Policy and Management, Tsinghua University, Beijing, China.

HE Xiaopei, Chengdu Government Office, Chengdu, China.

## ABSTRACT

*With the development of Internet of things (IoT), there are many smart city projects for improving urban services recent years. Community is a basic subunit of a city, smart community involves those important affairs which most directly related to unban services in a smart city overall planning. Based on the dualism of information architecture and business model by Kuk and Janssen (2011), the study proposed a theoretical framework to understanding the relationship among IoT and smart community which regards a path of smart community development as a configurations set including both information architecture factors and business model patterns. To explore successful configurations based on the proposed framework, we select 69 communities from Beijing, China and deal with the case material of those communities by using the method of coding and scheme matching before analysis. Then, using the Qualitative Comparative Analysis (QCA) method, we found that a successful smart community depends on the integration between information architectures and business models, and different business model patterns rely on different information architecture factors.*

## 1. Introduction

The Internet of Things (IoT) is the connection of objects via the Internet from the physical world that are equipped with sensors, actuators and communication technology (Djkman et al. 2015). Since the IoT is considered as a representative technological innovation for future computing and communications, more than eighty percent of experts predict IoT will have widespread and beneficial effects on the everyday lives of the public by 2025 according to an expert survey conducted by PEW (Pew Research Center, 2014). In particular, survey respondents expect the IoT to be evident in communities: Embedded devices and smartphone apps will enable more efficient transportation and give readouts on pollution levels. "Smart systems" might deliver electricity and water more efficiently and warn about infrastructure problems (Pew Research Center, 2014). IoT is more than smart homes and connected appliances. It scales up to include smart cities (Kobie 2015). The availability of different types of data, collected by a pervasive IoT, may also be exploited to increase the transparency and promote the actions of the local government toward the citizens, enhance the awareness of people about the status of their city, stimulate the active participation of the citizens in the management of public administration, and also stimulate the creation of new services upon those provided by the IoT (Zanella et al. 2014). Therefore, IoT has been considered an important technological infrastructure for fulfilling smarter lives of human beings.

Mahizhnan (1999) puts forward the concept of smart city in which enhancement of urban residents' life quality serves as a development goal. In 2008, the IBM Corporation's proposal of the concept of "Smart Planet" unlocks the door of practical exploration into the smart city, and it enables the concept to spread up worldwide (IBV 2009). With the development of technologies such as Internet of things, cloud computing, and big data, city development in China transit from a digital city stage to a smart city stage. To promote orderly and healthy development of smart city, many cities listed "construction of a smart city" into their 12th Five-Year Plan Outline (2011-2015), including Beijing, Shanghai, Nanjing, Guangzhou, Ningbo, and Shenzhen (Wang et al. 2012). In December 2012, Ministry of Housing and Urban-Rural Development published a list of 90 pilot cities, and further published a new list adding another 103 pilot cities in August 2013. National Development and Reform Commission (NDRC) also promulgated the guidelines for promoting the healthy development of smart city in August 2014 (NDRC 2014). It can be predicted that under the guidance of policies, a new round of smart city construction is upcoming during the 13th Five-Year Plan (2016-2020).

A community is a basic unit and cell of a city a. Many dimensions of development goals of a smart city are needed to be implemented at the level of communities, and smart community construction is means for "landing" the smart city construction. With unceasing propulsion of government transition, social management becomes more onerous, and accordingly promoting community self-management and community construction becomes an urgent need for city management. The development of information technology provides favorable conditions for a community to enhance the level of self-management and service. Advanced technologies, for example, IoT, enables communities to integrate its various resources more effectively, promotes citizens' voluntary participation, and provides customized service, and thereby truly realizing community self-management. In practice, IoT infrastructure is largely deployed at the community level, and the link between IoT and smart community reflects the relationship between IoT and smart cities at a micro-level, which is beneficial for understanding the relations between IoT and smart city.

IoT and smart city are the two sides of a coin. Smart city aims to provide front-end services and applications, and IoT focuses on rear-end infrastructure. The two concepts are interchangeable in many occasions. Understanding the relationship between IoT and Smart city is conductive to city development in the networking information technology context. Comparing cases from two Holland cities, Kuk and Janssen (2011) discuss two development paths of smart city construction: business models and information architectures. They also demonstrate that the influence of different practice paths on the development level of smart city. Since IoT overlap with the concept of information architecture, Kuk and Janssen (2011)'s work maybe could be considered as a reference for understanding the relationship between IOT and smart city. However, compared with increasingly enriched practices of smart communities, the theoretical exploration and empirical studies on smart community development still remain limited.

Based on multi-case studies of the Chinese smart community construction, this study defines and refines the two types of configurations following Kuk & Janssen (2011)' work for understanding the relationship among IoT (information architectures) and smart city (business models) in the urban communities context. The research questions are: (1) how to select an effective smart community practice path when resources are limited? How to combine those critical factors in two practical configurations (business models and information architectures) to improve the development level of smart community? Employing Qualitative Comparative Analysis (QCA) method, the analysis and discussion are based on 69 cases pilot smart communities from Chaoyang District, Beijing. The implications of results are also discussed in the concluding remarks.

## 2. Literature review

### 2.1 Internet of Things: Key Technologies and Networking Values

IoT enables new forms of communication between people and things, and between things themselves: from anytime, any place connectivity for anyone, we will now have connectivity for anything (International Telecommunication Union, 2005).The IoT is initiated by the use of Radio-Frequency Identification (RFID) technology, which is increasingly used in logistics, pharmaceutical production, retail, and diverse industries (Li, Xu, & Zhao, 2015). The emerging wirelessly sensory technologies have significantly extended the sensory capabilities of devices and the IoT represents the next generation of Internet, where the physical objects could be accessed and identified through the Internet. (Li et al., 2015).

The basic idea of IoT is the pervasive presence around us of a variety of things or objects, such as RFID tags, sensors, actuators, and mobile phones, which are able to interact with each other and cooperate with their neighbors to reach common goals (Atzori, Iera, & Morabito, 2010). Near Field Communications (NFC) and Wireless Sensor and Actuator Networks (WSAN) together with RFID are recognized as "the atomic components that will link the real world with the digital world" (Atzori et al., 2010, p. 2789).

The expansion of big data and the evolution of IoT technologies play an important role in the feasibility of smart city initiatives. Big data offer the potential for cities to obtain valuable insights from a large amount of data collected through various sources, and the IoT allows the integration of sensors, radio-frequency identification, and blue tooth in the real-world environment using highly networked services (Hashem et al., 2016). In addition, IoT can benefit from the unlimited capabilities and resources of cloud computing. Also, when coupled with IoT, cloud computing can in turn deal with real world things in a more distributed and dynamic manner (Zheng, Martin, Brohman, & Xu, 2014).

Although IoT is derived from sensing technologies, its growth in networking value is undoubtedly dependent on network connectivity, as well as cloud computing and big data on this basis.

**2.2 Smart City: Success Factors and Popular Fields**

As a product of information technology and urbanization, smart city provides a new perspective for urban management and innovation. Hollands (2008) contends that "one of the key elements which stands out in the smart (intelligent) city literature is the utilization of networked infrastructures to improve economic and political efficiency and enable social, cultural and urban development" (p. 307). Caraliu, Bo & Nijkamp (2011) argue that a city to be smart "when investments in human and social capital and traditional (transport) and modern (ICT) communication infrastructure fuel sustainable economic growth and a high quality of life, with a wise management of natural resources, through participatory governance" (p. 70). Smart city services are increasingly becoming "a norm rather than exception in developing and managing city services for citizens" (Lee & Lee, 2014, p. 93). The concept of smart community emerges under the background of smart city. Community is a basic functional unit of a city, and the development goals of smart cities need to be implemented and realized at the community level.

The Centre of Regional Science at the Vienna University of Technology identified six main components of a smart city: a smart economy, smart people, smart governance, smart mobility, a smart environment, and smart living (Giffinger et al. 2007). These six components are based on theories of regional competitiveness, human and social capital, civil participation, transport and ICT economics, natural resources, and quality of life (Giffinger et al. 2007; Caragliu , Bo & Nijkamp 2011). In addition, the Intelligent Community Forum defined critical success factors for the creation of Intelligent Communities: Broadband, knowledge workforce, innovation, digital equality, sustainability, and advocacy. Each year, the Intelligent Community Forum presents an awards program for intelligent communities that have taken a leadership role in promoting Internet technology and applications.

Linking key actors (universities, industry, government, and civil society) to the above-mentioned main dimensions of a smart city (smart economy, smart mobility, smart environment, smart people, smart living and smart governance), Lombardi et al. (2012) proposed five clusters of smart city performance indicators: smart governance (related to participation); smart human capital (related to people); smart environment (related to natural resources); smart living (related to the quality of life); and smart economy (related to competitiveness). In summary, these studies all suggest develop smart economy and smart governance based on smart people, thereby efficiently utilizing resources and achieving smart life (Kourtit & Nijkamp 2012, Mahizhnan 1999).

Moreover, many scholars contended that smart cities need to be designed to improve accessibility and allow the inclusion of all kinds of citizens (Mora, Gilart-Iglesias, Pérez-del Hoyo, & Andújar-Montoya, 2017; Rashid, Melià-Seguí, Pous, & Peig, 2017). For instance, motor disabled people like wheelchair users may have problems to interact with the city. IoT technologies provide the tools to include all citizens in the Smart City context (Rashid et al., 2017).

IoT is already delivering benefits to cities like Los Angeles and Oslo, which have experienced energy savings of more than 60 percent by moving to smart street lighting. Other cities have seen similar significant savings by deploying smart waste management solutions, reducing $CO_2$ emissions, and increasing citizen satisfaction through smart parking and traffic management (Zanella, Bui, Castellani, Vangelista, & Zorzi, 2014). In the area of health, digital health has allowed for patients to stay less in hospitals and more at home, thanks to monitoring sensors and devices, transmitting alerts and other relevant data over mobile networks and the Internet (Kamel Boulos & Al-Shorbaji, 2014). Kamel Boulos and Al-Shorbaji (2014) argue that IoT-powered smart cities stand better chances of becoming World Health Organization's healthier cities. With regard to education, mobile computing devices and the use of social media created opportunities for interaction and collaboration, and allowed students to engage in content creation and communication using social media and Web 2.0 tools with the assistance of constant connectivity (Gikas & Grant, 2013).

**2.3 IoT vs Smart City: Information Architectures vs Business Model**

Many regions and countries in the world have developed smart city policies and projects, such as Europe Union's "Living Lab", British "Smart Bay" Project, "I-Japan Strategy 2015", and "U-Korea Development Strategy". Although smart city and community projects have different focuses, these projects are developed mainly through two approaches: information architectures and business models (Kuk & Janssen, 2011). Comparing two cities in the Netherlands, Kuk and Janssen (2011) contend that there are two ways cities acquire the smart city status: information architectures and business models. While the approach of business models is to improve and increase front-end services to achieve rapid accumulation of commercial value, the approach of information architectures primarily aims to improve rear-end services for processing information with low cost and high efficiency. The first case accumulated business value faster with more new services made available to the public. In contrast, the second case was more resource-intensive and relatively slower in bringing new services to the general public, but the services were much improved and sustainable over time (Kuk & Janssen, 2011).

The approach of information architectures emphasizes that IT infrastructure is the foundation of smart city and community construction (Bartlett et al. 2011, Gann et al. 2011). It generally integrates a variety of information service resources and establishes unified data and service platform to make each department and system connected. On the other hand, the approach of business models focuses on the role of human being. It believes that the construction of smart cities is not all about investing and upgrading infrastructures, but exploring and utilizing human's wisdom to improve urban management. The approach of business models takes full advantage of ICT technology and public data, and provide one-stop service to enhance the value of service provision and political participation (IBV 2009).

Any realistic understanding of what it means to be a smart city needs to specify the type of business models being used and ensure that the information architecture is able to support the desired business models (Kuk & Janssen, 2011). In the context of smart city, IoT is a key enabler in the transformation towards smart cities (Kamel Boulos & Al-Shorbaji, 2014). That is, IoT is a major element of the information architecture, and smart city construction is one of the business models.

In short, the creation of smart communities still lies in an initial phase. Although countries have started the smart community construction through the approaches of information architectures and business models, there isn't a unified and effective development path of smart community. Based on pilot smart communities in China, this study employs empirical methods to analyze the influence of information architectures and business models on development level of smart communities. Moreover, the relationship among sub-factors of information architectures and business models will also be explored.

## 3. Theoretical framework and research hypotheses

Public sector managers should have a clear idea what it is they hope to achieve with limited resources at their disposal and select appropriate transformation strategy (Kuk & Janssen, 2011). There is no lack of discussion about the attentions of nations, organizations or their leaders in both economics and politics (Davenport & Beck 2000; Jones & Baumgarther 2005; Wood & Peake 1998). Different from the psychological perspective of attention studies on the individual level, the research focuses in public administration are around the relationship among the different types of resources and attentions (Ocasio 1997). Rational governors tend to process a few issues in a serial rather than parallel fashion based on their resource views, attending to some issues before moving on to others (Jones 1994). Therefore, the different schemes to prioritize issues imply the different choice of strategies. The work also follow the logic of limited resources when discuss the more successful strategies of smart city or smart community projects.

### 3.1 Relationship among Information Architectures, Business Models and Smart Community Development

Business models include governments' activities such as providing personalized services, developing smart applications, and expanding service types by virtue of information infrastructure and intelligent platform and the like. Information architectures refer to construction of forms of information organization and presentation by governments' increase of infrastructure construction investment, integration of information platform, and development of unified database. Kuk and Janssen (2011) argue that the two approaches (business models and information architectures) are not mutually exclusive and should be considered complementary. New business models and supporting information architectures should be developed in parallel with an approach involving both the front-end and the back-end. Therefore, this study argues that the construction of the two aspects-information architectures and business models-is not separated completely, but shall be integrated with each other during the process of smart community construction. Accordingly, the first research hypothesis of this study is:

*Hypothesis H1: A successful smart community relies on path integration between information architectures and business models.*

### 3.2 Information Architectures Based on IoT

Based on prior research, this study proposes five sub-factors of information architectures, including networking, data warehouse, terminals, sensors, interaction & payment. We discuss the detail of those five factors as follow:

- **Networking:** The most principal feature of IoT is connection. Both physical world and virtual world need a strong network to connect (Djkman et al. 2015). Networking, known as one of the IoT factors, are computer networking devices which required for communication and interaction among devices both on Internet and local area network. Networking hardware often includes devices for digital TV network, broadband communication network, wireless network, and some switching network among them in the smart city and smart community context.

- **Data warehouse:** The expansion of big data and the evolution of IoT have played an important role in the feasibility of smart city initiatives (Hashem et al. 2016). Storage, management and mining of data from the physical world are basic work in the IoT. Data warehouse, considered as an updated database system for better extracting, transforming and loading, includes several database sub-systems of population data, social media data, administrative data, etc. in the smart city and smart community context.

- **Terminals:** In the broadest definition, every node on the IoT could be called a terminal including RFID and many types of sensors. In this paper, however, terminals, considered as ones following traditional computer terminal definition, are electronic hardware devices that is used for both entering and displaying information. In the smart city and smart community context, terminals refer to computers, mobile phones, tablets, public displays etc.

- **Sensors:** Sensor is a device, module, or subsystem whose purpose is to detect events or changes in its environment and send the information to other terminals. As entrances to physical world, billions of sensors are the basis of the world of IoT (Yan et al. 2015). With the public's demand for living environment improved, air quality monitoring fire smoke detection, high concentration gas alarm, toxic and hazardous substances monitoring, and some other sensors have been more often considered in the smart city plans in China (Zhang et al. 2015).

- **Interaction & Payment:** The potential of IoT lies in the interaction among nodes working together toward value co-creation (Ghangbari et al. 2017). While the core spirit of Internet changes from Web 1.0 to Web 2.0, the interactive functions in terminals have been paid more attention. In the smart city and smart community context, some interactive functions, especially payment functions of financial technologies are both popular and important to citizens.

### 3.3 Business Models toward Smart City

Based on prior research, this study proposes five variables of business models, including public information, facilities management, healthcare service, education service, and accessibility service. We discuss the detail of those five variables as follow:

- **Public Information:** Janssen, Kuk, and Wagenaar (2008) contend that one of the e-government business models is "content provider", that is, governments provide static and dynamic content, including contact information, organization information, product and service information, and news. Based on Internet and information system distribution channels, smart community is able to provide information related to public policy, safety, and job, improving government transparency and enhancing interaction between community managers and residents.

- **Facilities management**: To improve the level of community management and service, smart communities provide online service platform for local residents. Based on technologies such as IoT and cloud computing, smart community can realize remote home appliance control through wireless terminal equipment. In addition, with intelligent parking system and real-time information, residents in smart communities could complete the process of locating parking spaces, parking their cars and retrieving them by accessing mobile phone apps.

- **Healthcare service**: Cities around the world could benefit from, and harness the power of, IoT to improve the health and well-being of their local populations (Kamel Boulos & Al-Shorbaji, 2014). Smart community construction can improve community health service level through establishing electronic medical records system and providing online appointment system and health consultation for community residents.

- **Education service**: Many smart city initiatives list education as one of the most important areas (Díaz-Díaz, Muñoz, & Pérez-González, 2017). Using IT technologies and means, smart communities build lifelong online learning platform, distance education platform and Internet library; and provide Internet skills training and professional training for community residents.

- **Accessibility service**: Some individuals are vulnerable because the nature and/or severity of their illnesses or disabilities create special challenges in obtaining the needed range of services. Communities play a significant role in providing services to meet the specific needs of vulnerable populations. Based on latest communication and positioning technologies for smart sensing, such as Augmented Reality (AR) and RFID technologies, smart cities and communities are able allow the inclusion of all kinds of citizens (Mora, Gilart-Iglesias, Pérez-del Hoyo, & Andújar-Montoya, 2017; Rashid, Melià-Seguí, Pous, & Peig, 2017).

**3.4 Different Influences of Information Architecture Factors on Business Model Patterns**

Sophisticated information infrastructure constitutes a precondition and foundation for smart management and service of communities, and thus the level of community information architecture would to a certain degree influence the development of business models. The creation of smart community raises different demands for information architecture sub-factors in terms of construction of business model sub-patterns. Accordingly, this paper presents the second research hypothesis.

*Hypothesis H2: Various business model patterns depend on different information architecture factors.*

Information architectures and business models are important constituent parts of the smart community construction. Based on current practices on the smart community construction, since each community is affected by financial resources and leader's awareness and so on, it is different in actual situations of construction for five aspects of information architectures. These variance combinations would cause different influences on the construction of business model sub-patterns, e.g., promoting, suppressing, or without remarkable effect. Thus, this study deems different combinations of information architecture sub-factors as specific configurations for forming business model sub-patterns, and conducts explorative researches based on verification of the two hypotheses to probe into specific information architecture factors on which various business model patterns rely, thereby searching effective configurations for development of the smart community construction. The framework is presented in Figure 1.

-------------------------------------------

Insert Figure 1 about here.

-------------------------------------------

## 4. Research method

The data used in this paper come from Information Network Center of Chaoyang District, which is a principal department responsible for smart community construction in Chaoyang District. The data include various texts, pictures, and video files for declaring starred smart community by 69 communities from Chaoyang Area.

Since respective factors in different practice paths of smart community construction are of combination relation in effect, it is difficult to find out combinational logic among different factors with use of statistical methods, such as regression, etc. Additionally, small sample confronted with the research also makes the effect of conventional statistical analysis method restricted, and hence a qualitative comparative analysis method is selected in this paper.

This study applies the cases of 69 communities as units of analysis, and implements quantification of textual contents by use of the method of coding and scheme matching techniques (Guo et al. 2010, Reimers & Johnston 2008). The related research procedures experience cross check of many researchers (Yin 2003). Material from 69 communities is first reorganized; key information is extracted to form a text database; the text database is classified and reorganized and encoded according to 10 factors of information architectures and business models. Meanwhile, the smart community development level in this paper is measured by scores of an official performance evaluation of communities conducted by Chaoyang local government based on a communities satisfaction survey. The scores of performance appraisal of the smart community are obtained from researchers' depth interview and collection of score questionnaire from many specialists of the principal departments of the smart community in Chaoyang District.

## 4.1 Valuation Principle and Weight

In order to avoid subjective judgment and discretion, this paper formulates unified valuation principle and standard for evaluation, and endows scores and weight to text data which are classified and arranged. Subsequently, a number of specialists are invited to score the reorganized community text material as a function of the valuation principle and revise the scores with great difference for multiple times so as to eventually obtain the scores of respective analysis variables in each community.

The unified valuation principle for all indexes include: (1) valuation field of all indexes is set as an interval [0, 10], and grades are scored from 0, the worst, to 10, the best. (2) As for the valuation principle employed for the index which is directly determined by "yes" or "none",

"none" corresponds to 1, and "yes" corresponds to 10. (3) If there is merely qualitative material, the start score for the index is 1; if there is merely quantitative material, the start score for the index is 3; if quantitative material and qualitative material appear simultaneously, the start score for the index is 5. (4) All evaluation indexes are positive indexes, and the valuation level represents advantages and disadvantages of the smart community construction. (5) All secondary indexes have equal weight in this paper.

### 4.1.1 Scoring Principle for Quantitative Index

Based on the above scoring principle, this paper intends to use a deviation method to set estimation scale, and assignment is performed as a function of data value of a single index, i.e., calculating all numerical mean value (standard value) and standard deviation of the single index. The bonus points of 1 to 5 are determined from the multiple relation between data value and standard deviation (see Table 1), and the summed scores are no more than 10.

-------------------------------------------

Insert Table 1 about here.

-------------------------------------------

For example, upon indexes of hardware infrastructure-resident-specific computer, there are totally 31 numerical data whose average is 8.26 and standard deviation is 7.02. By taking the average 8.26 as the standard deviation in combination with actual conditions of 69 communities, estimation scale scores are set with more than 0.5 times and 1 times standard deviation and less than 0.5 times and 1 times standard deviation respectively (see Table 2).

-------------------------------------------

Insert Table 2 about here.

-------------------------------------------

**4.1.2 Scoring Principle for Qualitative Index**

A large proportion of the material from 69 communities arranged in this paper is qualitative. In order to facilitate quantifying the qualitative material, this paper endows marks according to two dimensions, i.e., clarity degree and execution level of the index.

(1) The clarity degree of the single index becomes gradually clear with the promotion of the smart community construction. Various clarity degrees are divided into a fuzzy phase and distinct phase in this paper (see Table 3).

-------------------------------------------

Insert Table 3 about here.

-------------------------------------------

(2) Execution level endows marks according to actual implementation of the index to indicate the extent to which the index is executed. According to the execution level of the index, the material of the smart community construction is divided into low execution phase and high execution phase. The valuation is assigned with 1 mark in the low execution phase, and the high execution phase is endowed with 1 mark for an increase of 1-2 items each based on refined execution projects upon the single index, with the summed scores no more than 10.

In accordance with unified scoring principle for respective indexes, the paper formulates specific score and grade standards for each evaluation index so as to facilitate consistency of many people's assignment in the next research. The material of 69 communities is completely quantified through the assignment steps.

## 4.2 Principle and Application of QCA Method

Firstly proposed by Ragin (1987), Qualitative Comparative Analysis (QCA) method is a case study-oriented research method. With constant dialogue between theoretical and empirical material and two types of specific operational methods of a crisp set and fuzzy set, the QCA method analyzes causal relationships between respective research topics in sample data, and it is an effective method for analyzing middle and small sample data (Ragin 1987, 2000). The QCA method is not an alternative to quantitative analysis method, and they have different analytical logic. The QCA assumes that the logical relationship between explanatory variables and explained variables are non-linear, and the effect thereof is interdependent. Furthermore, there are potentially a variety of different cause combinations which bring about occurrence of the same social phenomenon. The factors which influence practice configurations of the smart community in this research are multi-conditional and concurrent. The unit of analysis of the qualitative comparative analysis method is not a single case but various condition combinations, and it can find out all possible logical combinations between explanatory variables and explained variables.

On specific operational level, research case and variable shall be first determined when using the QCA method. In this paper, for one thing, 10 indexes of information architectures and business models serve as explanatory variables during verification of H1, and the smart community development level serves as explained variables to study different influences of information architectures and business models on the smart community development level. For another, during verification of H2, five factors of information architectures serve as explanatory variables, and five patterns of business models serve as explained variables to study influences of different information architecture sub-factors on business model patterns.

Next, a membership degree, i.e., scores of 0-1, of the explanatory variables and explained variables in the case is determined from observation of the case. Since the data of fuzzy set QCA must be between 0-1, the calibrate functions (x, n1, n2, n3) which are provided based on the QCA method correct the original scores (Ragin 1987, 2000), i.e., adding qualitative narration in complete dichotomy of 0 and 1 to distinguish the degree of variables. All data after conversion are in the 0-1 set.

Afterward, a truth table is constructed based on membership degree and fuzzy set of explanatory variables and explained variables. If paradoxical combinations appear in the truth table, i.e., the situation where the same variable combinations cause different results, it is required to recheck the variable data of the paradoxical combinations, or to directly exclude the paradoxical conditions in case of sufficient argument.

**4.3 QCA Operation and Hypothesis Testing**

On this basis, analysis software of Fs QCA 2.5 is utilized to select a fuzzy set for operation. The computational results are represented by three kinds of solutions, i.e., complex solution, parsimonious solution, and intermediate solution, wherein the parsimonious solution typically explains causal variables in a simplest manner, which significantly differs from the other two solutions. Hence, this paper uses the complex solution and intermediate solution to explain the relations between variables.

In order to validate whether QCA's operation of solutions is effective, there is a need to substitute association of variables provided by QCA back the original case for verification. If there is a case which supports and runs solutions, it is proved that the solutions are effective; otherwise the solutions will be eliminated.

Lastly, research hypotheses are validated by efficient solutions. Meanwhile, this paper proposes null hypothesis and alternative hypothesis of H1 and H2 under such an empirical context. During verification of H1, respective information architecture factors and business model patterns are used as explanatory variables, and the smart community development level is used as explained variables. During verification of H2, respective information architecture factors are applied as explanatory variables, and various business model patterns are applied as explained variables. It is derived from verification of null hypothesis and alternative hypothesis to which subset of the explanatory association of variables the explained variables belong, and influence factor combinations for the studied issues are refined.

**4.4 Validity and Reliability**

In order to guarantee validity of research, this paper first uses a plurality of channels to obtain material during data collection and sorting phase (Yin 2003). The primary evidence originates from channels, such as case record, interview, direct observation and physical evidence, etc. Next, in terms of evidence chains, it is assured that there is explicit association between the studied issues, collected material and conclusions. In this paper, the hypothetical proposition which can be derived by cases is first analyzed via the QCA, and the proposition is then substituted back the original case to find out the specific case material which can support such a conclusion.

In the aspect of ensuring research reliability, on the one hand, key information is extracted from the collected material before classifying and encoding the material to form a text database which provides readers with original material for independent validation. On the other, this paper formulates specific valuation principle and standard and grades for multiple rounds, thereby preventing subjective judgment and discretion from affecting accuracy of results (Yin 2003).

Moreover, we also report solution consistency indexes and solution coverage indexes in results demonstration by following Fiss (2011)'s work.

## 5. Results and discussion

Although it was found from depth interview for confirming sub-factors of information architectures and business models that various communities employ differentiated practical paths during the smart community construction, the analysis results by using fs QCA 2.5 show clearer picture of rich configuration patterns among different communities.

### 5.1 Information Architectures, Business Models and the Smart Community Development

Five sub-factors of information architectures and five sub-patterns of business models are selected as explanatory variables, and the level of smart community development is used as explained variables. The analysis software fs QCA 2.5 is employed to conduct standard analyses. Following Fiss (2011)'s recent work on Academy of Management Journal, we demonstrate our results by using symbols rather than mathematical formulas. As referring to the results, it is found that the results of complex solution and intermediate solution are consistent, and thus those two solutions are combined for explanation together (see Table 4). In analysis of QCA results, value of unique coverage represents which of the corresponding combinations can better impact the smart community development level. In view of the entire solution coverage and solution consistency, the coverage is 0.33, and it may be restricted by a sample size, which illustrates that there are possibly other factors to influence the smart community development level. However, the whole solution consistency scores up to 0.93, illustrating that information architecture and business models can better explain the level of the smart community development.

-------------------------------------------

Insert Table 4 about here.

-------------------------------------------

The conditional configurations which can be observed from the table 4. Terminals and Sensors seem more important IoT infrastructures to smart communities in information architecture level while a data warehouse system is not indispensable to community level. Although business models present differences between two configurations, the accessibility service is emphasized by neither A nor B. Considering the minority is benefit from accessibility, the result is not hard to understand. However, we maybe concluded the future community evaluation indexes should be adjusted for adapting accessibility requirement, rather than accessibility is not important.

In order to support association of variables provided by QCA, the association of variables corresponds to sample of case for verification and explanation. It was found from observation that two configurations approximately correspond to one sample of community respectively, i.e., Jianwai Community and Maizidian Community respectively. Jianwai community is located in the city center of Beijing, with good experiences in public information release from several years ago. In recent years, the investment of Jianwai community on IoT focus on a variety of terminal construction. Especially the housebound convenient payment has become one of highlights of the community. It also provides the diversity of the extended community services. Different from Jianwai community, the Maizidian community is far away from downtown, then has a different smart community development configurations. High-speed networking environment construction investment is still needed, while the community healthcare and education services also have a wider range of demand. In total, It can be derived from original formulas and corresponding case number combinations that the of variables corresponding to the smart community with high performance must include information architecture factors and business model patterns simultaneously, and a successful smart community depends on the configuration integration between information architectures and business models, that is, the hypothesis H1 is validated.

**5.2 Relationship between Information Architecture and Business Model**

Five factors of information architectures are selected as explanatory variables, and five patterns of business models are used as explained variables to establish a truth table respectively, perform standard analysis, and incorporate complex solution with intermediate solution, thereby obtaining factor combinations of information architectures which implement various business model patterns (see Table 5).

------------------------------------------

Insert Table 5 about here.

------------------------------------------

It can be seen from the Table that there is difference between factor combinations of information architectures which correspond to various business model patterns, that is, the research hypothesis H2 is validated. On this basis, this paper conducts explorative researches by probing into information architecture factors on which various business model patterns specifically depend, thereby searching an effective configurations for the construction of the smart community. For this target, five factors of information architectures are selected as explanatory variables, and five patterns of business models are used as explained variables respectively for QCA analysis.

With regard to healthcare service of business models, there are have eight combinations includes four three-factor combinations and four four-factor combinations of information architectures with higher frequency of occurrences. It can be known from further analysis on those combinations that the healthcare service more relied on broad perspective terminal configurations, including terminals, interaction & payment, than the other ones. We consider that the community as a grass root node of e-the health system, is mainly responsible for the fulfilling of healthcare information collection, display, and providing the necessary interaction and payment channels. To clarify the understanding as mentioned above of community healthcare service development is important.

For the other three, Public information facilities management and education, we also discuss briefly as follow. In terms of public information service, combination of networking, data warehouse and terminals play the important roles of boosting the improvement of public information service. However, the sensors, interaction and payment do not perform the remarkable function yet in promoting the improvement of information management and service. Regarding of facilities management, just one conditional combination is supplied by the QCA, which stresses the presence of networking, data warehouse and sensors. Meanwhile, when sensors and pyment are absent or the construction level thereof is low, such a factor combination plays the role of boosting construction and management service level of communities. There are also three kinds of information architecture factor combinations with more corresponding cases for education business models. It is found from observation that terminals exists simultaneously in three associations of variable, and thus, it may be regarded as an essential condition for education service. Further, networking also performs the vital function in promoting the improvement of the education service.

Last but not least, for accessibility issues, there are two conditional combinations which result in favorable accessibility service, wherein interaction & payment is a prerequisite for special crowd accessible service of the smart community. There is interesting to note that, however, in the rest of the four factors, two combinations seem to be the opposite to each other. The results probably reflect the two thoughts for improving accessibility in smart city projects: The first path is developing the networking environment, and solving the challenges based on the external resources via Internet, while the second is developing local terminals and other facilities, and solving the challenges based on the internal capacities. For different contexts of the digital divide, such as for senior people, or for several types of disabilities, the applicability of the two strategies remains to be further discussed.

## 6. Concluding remarks

Based on the mentioned results, firstly, the smart community development level is jointly influenced by information architectures and business models, and a successful smart community depends on path integration between information architectures and business models. The results also could be considered as a validation of relationship between IoT and Smart Cities in the community level. Facing to the emergence of new technologies, such as IoT, cloud computing, big data, artificial intelligence, urban development obviously have more opportunities. The implications from the study address that, however, we should never forget that the opportunities could be turned into realities only when the technological characteristics integrated with urban service requirements, the business model of urban management.

Secondly, in a limited resources context, if finding prioritize issues for smart cities and smart communities is important, The work also propose a method to explore successful strategies of smart city or smart community projects. According to QCA results, we have found that various business model patterns rely on different information architecture factors. Therefore, during the construction of a smart community, on the one hand, emphasis shall be laid on supplementation and coordinate propulsion of informatization infrastructure construction and business model expansions. On the other, a community should combine with its own reality, develop its own comparative advantage, specify construction goal, and select an effective differentiated development path, especially face to some high investment technologies such as IoT. In addition, networking, sensors, interaction & payment are information architecture factors which are depended more frequently by respective business model patterns, and they shall be highlighted in the development of a smart community.

The study still has some limitations. First, as mentioned in the prior sections, since the community development evaluation is based on citizen satisfaction survey, it obviously cause systematic bias to the attention of the majority of people demand, and get some results like accessibility is not important. Second, IoT technological features and smart city business models are still developing and increasing, taxonomy always seems to be a difficult issue. For example, artificial intelligence may also bring new differentiation to interactive methods. However, the study proposed a new perspective to understand the relationship between IoT and smart cities by extending Kuk and Janssen (2011)'s information architecture and business model dualism, and provide a preliminary method to find smart service success configurations in community level. In follow-up studies, we will through a wider range of empirical studies to make up the limitations mentioned above and explore more deep research issues around smart city and smart community based on IoT.

**Acknowledgement**

This work was partially supported by the National Natural Science Foundation of China [71473143, 91646103], the Beijing Social Science Foundation [15JGA008].

**Table 1**

**Assignment Principle of Quantitative Index**

| Less than 1 times standard deviation | Less than 0.5 times standard deviation | Standard Value | More than 0.5 times standard deviation | More than 1 times standard deviation |
|---|---|---|---|---|
| +1 | +2 | +3 | +4 | +5 |

**Table 2**

**Scoring Principle of Resident-specific Computer Index**

| Grade | Number of Computers for Residents' Exclusive Use | Reference Assignment |
|---|---|---|
| More than 1 times standard deviation | 15.28 | +5 |
| More than 0.5 times standard deviation | 11.77 | +4 |
| Standard Value | 8.26 | +3 |
| Less than 0.5 times standard value | 4.75 | +2 |

| Less than 1 times standard deviation | 1.24 | +1 |

**Table 3**

**Scoring Principle for Clarity Degree of Qualitative Index**

| Clarity Degree | Specific Performance | Scoring |
| --- | --- | --- |
| Fuzzy Phase | The community does not think deeply according to the index description in guidance standard for the smart community of Beijing city, or does not redefine the index. | +1 |
| Distinct Phase | The community specifies the standard indexes in view of its own actual situations and stores them with the material of language and character so as to facilitate instructing development of the smart community construction | +2 |

**Table 4**

**The Successful Smart Community Configurations**

| Configuration | A | B |
| --- | --- | --- |
| Networking | ⊗ | ● |

| Information Architectures | Data Warehouse | ⊗ | ⊗ |
|---|---|---|---|
| | Terminal | ● | ● |
| | Sensors | ● | ● |
| | Interaction & Payment | • | ⊗ |
| Business Models | Public Information | ● | ⊗ |
| | Facilities Management | • | ⊗ |
| | Healthcare | • | ● |
| | Education | ⊗ | ● |
| | Accessibility Service | ⊗ | ⊗ |
| Key Indexes | Consistency | 0.93 | 0.90 |
| | Raw Coverage | 0.25 | 0.20 |
| | Unique Coverage | 0.13 | 0.08 |
| | Overall Solution Consistency | 0.93 ||
| | Overall Solution Coverage | 0.33 ||

**Notes: Black circles indicate the presence of a condition, and circles with "X" indicate its absence. Large circles indicate core conditions; small ones, peripheral conditions. Blank spaces indicate "don't care."**

**Table 5**

**Factor Combinations of Information Architectures on which Various Business Model Patterns**

|  |  | Networking | Data Warehouse | Terminal | Sensors | Interaction & Payment |  |
|---|---|---|---|---|---|---|---|
| Public Information | P31 | ● |  | ● |  | ● | Solution Consistency: 0.91 Solution Coverage: 0.52 |
|  | P41 | ● | ● | ● | ⊗ |  |  |
|  | P42 | ● | ● |  | ⊗ | ⊗ |  |
|  | P43 | ● | ● | ⊗ |  | ⊗ |  |
|  | P44 |  | ⊗ | ● | ● | ⊗ |  |
|  | P45 | ⊗ | ⊗ |  | ⊗ | ● |  |
| Facilities Management | F51 | ● | ● | ⊗ | ● | ⊗ | Solution Consistency: 0.91 Solution Coverage: 0.29 |
| Healthcare | H31 | ● | ⊗ |  |  | ● | Solution Consistency: 0.85 |
|  | H32 |  |  | ● | ● | ● |  |
| | H33 | | Ⓧ | ● | ● | | Solution Coverage: 0.75 |
| | H34 | Ⓧ | | ● | ● | | |
| | H41 | ● | | Ⓧ | Ⓧ | ● | |
| | H42 | Ⓧ | ● | ● | | ● | |
| | H43 | Ⓧ | ● | | ● | ● | |
| | H44 | ● | ● | Ⓧ | ● | Ⓧ | |
| Education | E41 | ● | | ● | Ⓧ | Ⓧ | Solution Consistency: 0.94 Solution Coverage: 0.44 |
| | E42 | ● | Ⓧ | ● | ● | | |
| | E43 | | Ⓧ | ● | ● | Ⓧ | |
| Accessibility Service | A41 | | ● | ● | ● | ● | Solution Consistency: 0.90 Solution Coverage: 0.48 |
| | A51 | ● | Ⓧ | Ⓧ | Ⓧ | ● | |

**Notes: Black circles indicate the presence of a condition, and circles with "X" indicate its absence. Blank spaces indicate "don't care."**

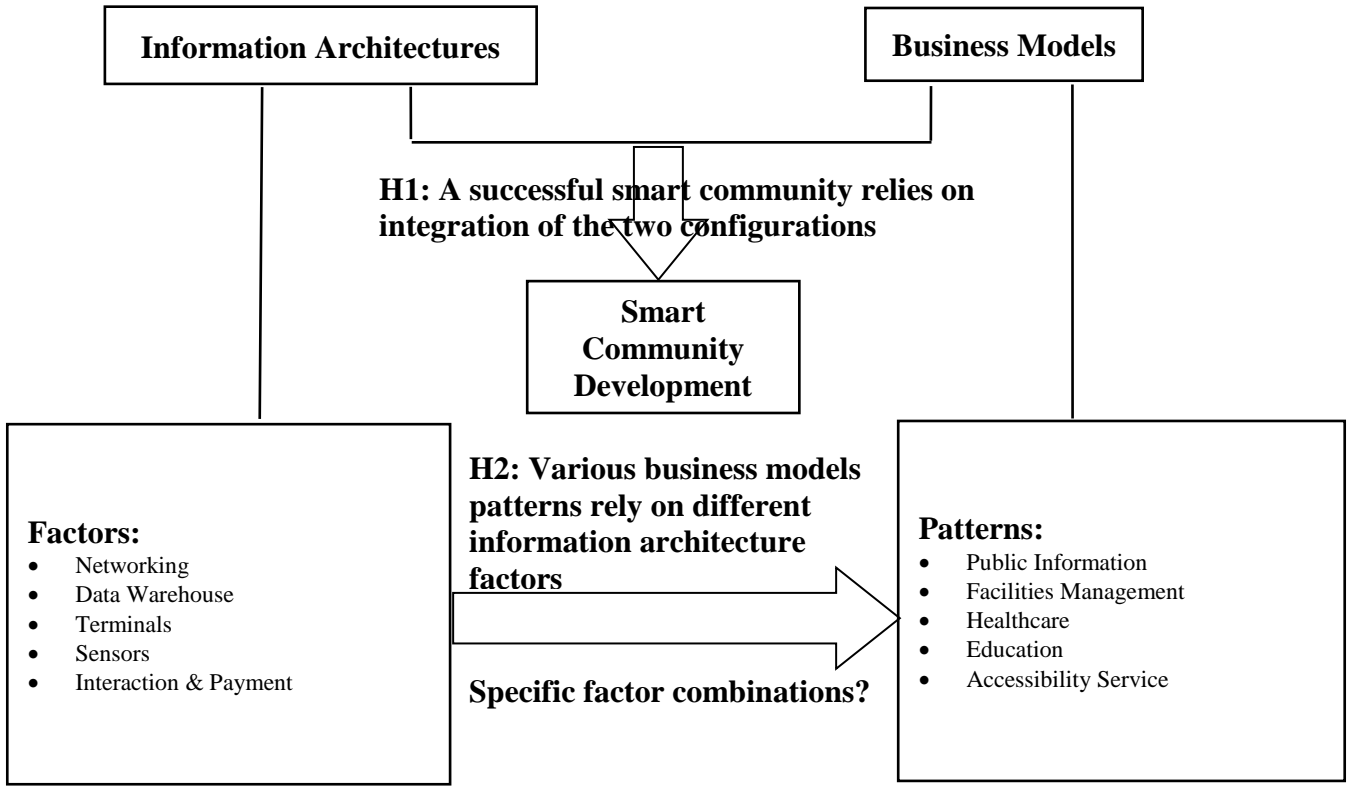

**Fig. 1. Research Theoretical Framework**